\documentclass{elsart1p}
\usepackage{graphicx}
\usepackage{wrapfig}
\usepackage{amssymb}

\begin{document}
\begin{frontmatter}
%
%
%
\title{Reaction plane dependence of neutral pion production in
  center-of-mass energy of 200~GeV Au+Au collisions at RHIC-PHENIX} 
%
%
\author{Yoki Aramaki, for the PHENIX Collaboration}
\ead{aramaki@cns.s.u-tokyo.ac.jp}
\address{Center for Nuclear Study, Graduate School of Science,
  University of Tokyo, 7-3-1 Hongo, Bunkyo, Tokyo 113-0033, Japan}
\begin{abstract}
  The integrated luminosity of RHIC-Year~2007 Au+Au run is 813~$\mu
  b^{-1}$, which is 3.5 times larger than that in RHIC-Year~2004 Au+Au
  run.   
  Additionally, a new detector was installed to determine reaction
  plane more precisely.  
  This detector is expected to provide better resolution for reaction
  plane determination.
  These advantages enable us to precisely measure the path length
  dependence of $\pi^{0}$ suppression and discuss parton energy
  loss mechanism more thoroughly. 
  We report the recent results for the reaction plane dependence of
  $\pi^{0}$ production.  
\end{abstract}
\end{frontmatter}
%
\section{Introduction}
It has been observed in central Au+Au collisions at Relativistic
Heavy Ion Collider (RHIC) that the yield of neutral pions at high
transverse momentum ($p_{T}$$>$5~GeV/$c$) is strongly suppressed
compared to the one expected from p+p collisions scaled by the
number of binary collisions.  
This suppression is considered to be due to the energy lost by hard
scattered partons in the medium (jet quenching), which results in a
decrease of the yield at a given $p_{T}$. 
Many theoretical models have been proposed to understand the parton
energy loss mechanism. 
For one thing, GLV method\cite{cite:GLV} as one of the calculations
predicts that the magnitude of energy loss is proportional to the path
length if the medium has static density.  
Studying the path length dependence of energy loss should help the
understanding of energy loss process. 

\section{Azimuthal angle dependence of $R_{AA}$}
\begin{figure}[htbp]
  \includegraphics[width=9cm]{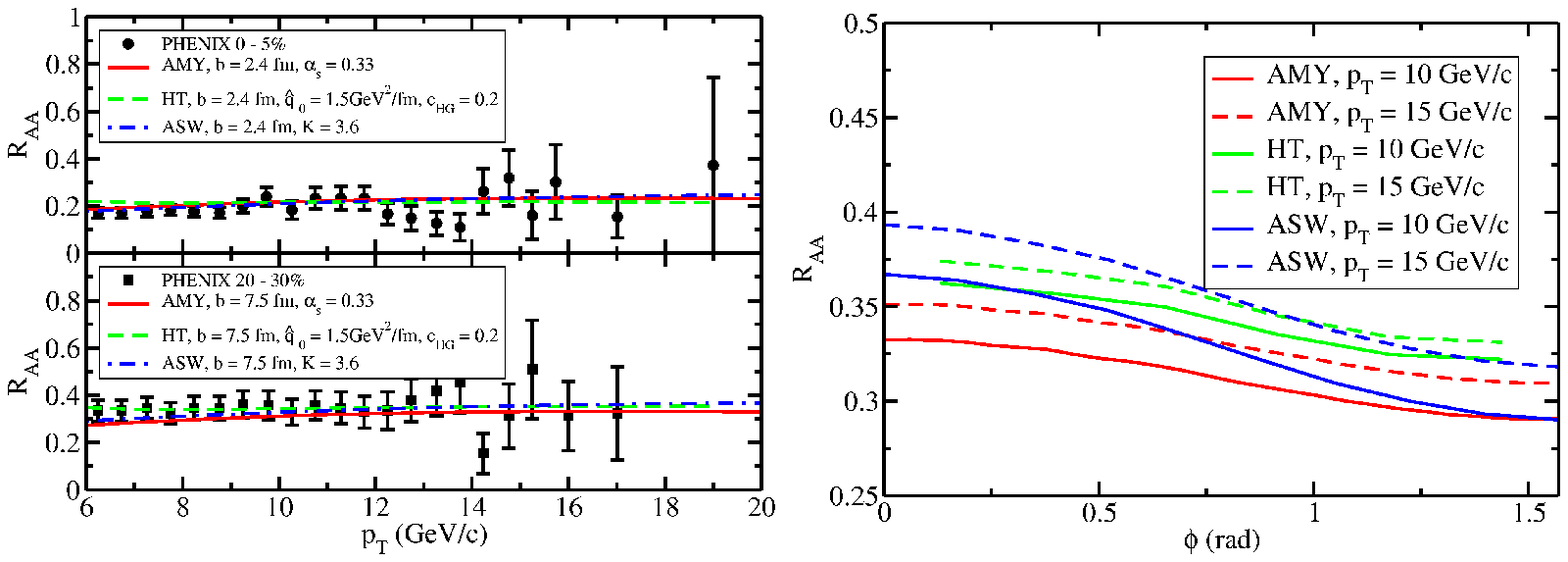}
  \includegraphics[width=4cm]{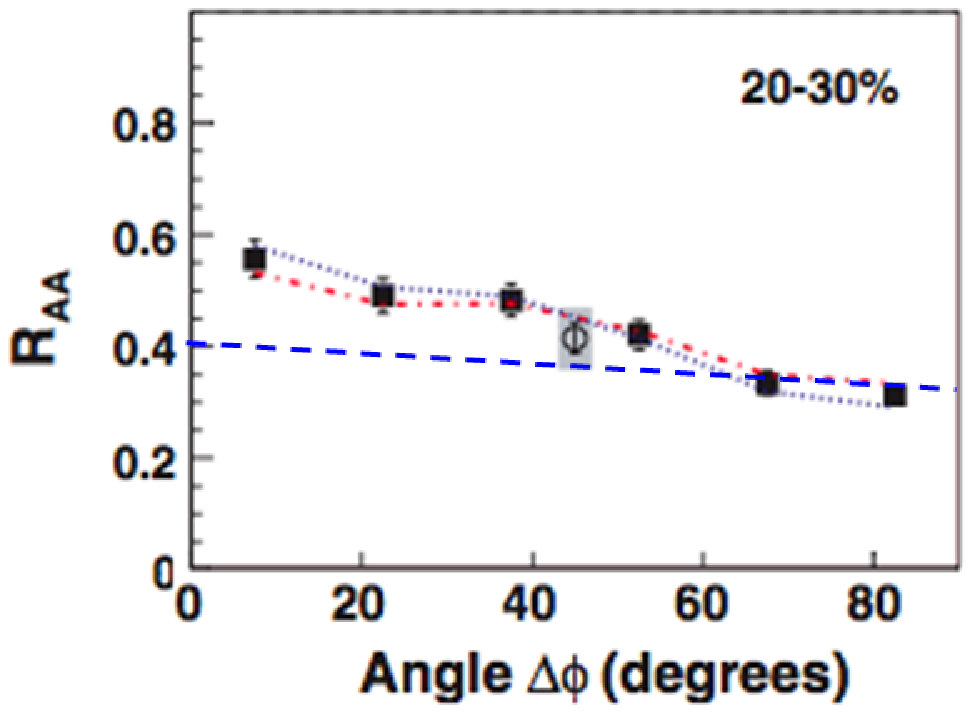}
  \caption{Left: $R_{AA}$ in Au+Au collisions at 0-5~$\%$ (top) and
      20-30~$\%$ (bottom) centrality calculated in the ASW, HT and AMY
      models compared with data from PHENIX\cite{cite:QM05pi0Raa}.
      Central: $R_{AA}$ as a function of azimuthal angle at $p_{T} =
      10$~GeV/$c$ (solid line) and $p_{T} = 15$~GeV/$c$ (dashed line)
      for all three models at 20-30~$\%$ centrality.
      Right: $R_{AA}$ as a function of azimuthal angle
      (5$<$$p_{T}$$<$8~GeV/$c$) at centrality 20-30$\%$ and blue
      dashed line is ASW model at $p_{T} = 15$~GeV/$c$.}
  \label{Fig:BassModelComp}
\end{figure}

Recently theoretical models (ASW\cite{cite:ASW}, HT\cite{cite:HT} and
AMY\cite{cite:AMY}) to describe parton energy loss mechanism which
involve the time-evolution of the medium produced at RHIC have
been proposed.  
These models succeeded in reproducing the centrality dependence of
$R_{AA}(p_{T})$.  
These theoretical curves and the preliminary data from PHENIX is shown
in Fig.~\ref{Fig:BassModelComp}\cite{cite:BassModelComp}.  
The central Fig.~\ref{Fig:BassModelComp} shows expected curves from
these models. 

As shown on the right panel of Fig.~\ref{Fig:BassModelComp}, these
models are still unable to reproduce the azimuthal angle dependence of
$R_{AA}$, even if we compare the curve with the largest variation, the
$p_{T} = 15$~GeV/$c$ of the ASW model to data at much lower $p_{T}$
(5$<$$p_{T}<$8~GeV/$c$).   

\begin{wrapfigure}[13]{r}{4.5cm}
  \includegraphics[width=4.5cm]{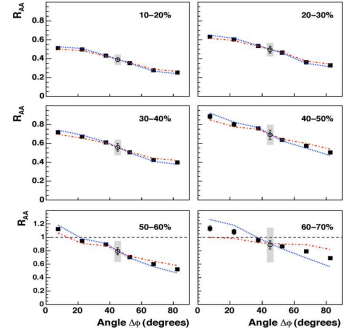}
  \caption{Azimuthal angle dependence of $R_{AA}$ integrated over
    3$<$$p_{T}$$<$5~GeV/$c$\cite{cite:pi0Raa}.}
  \label{Fig:pi0RaaDeltaPhi}
\end{wrapfigure}

When partons lose energy in the medium, the ratio of
$R_{AA}(\Delta\phi=0)$ to $R_{AA}(\Delta\phi=\pi/2)$ should be
different for each centrality bin.
Fig.~\ref{Fig:pi0RaaDeltaPhi} shows the azimuthal angle dependence of
$R_{AA}$ at 3$<$$p_{T}$$<$5~GeV/$c$.
As shown in Fig.~\ref{Fig:pi0RaaDeltaPhi},
$R_{AA}$(3$<$$p_{T}$$<$5~GeV/$c$, $\Delta\phi=0$) is about two times
larger than $R_{AA}$(3$<$$p_{T}$$<$5~GeV/$c$, $\Delta\phi=\pi/2$) for
all centrality bins.     
This may be in part be an effect of collective flow.
Since the influence of collective flow at high $p_{T}$ should be small,
we need to measure $R_{AA}(p_{T}, \Delta\phi)$ at higher $p_{T}$ to
minimize its effect.

\section{$v_{2}(\pi^{0})$ at high $p_{T}$}
\begin{figure}[htbp]
  \centering
   \includegraphics[width=11cm]{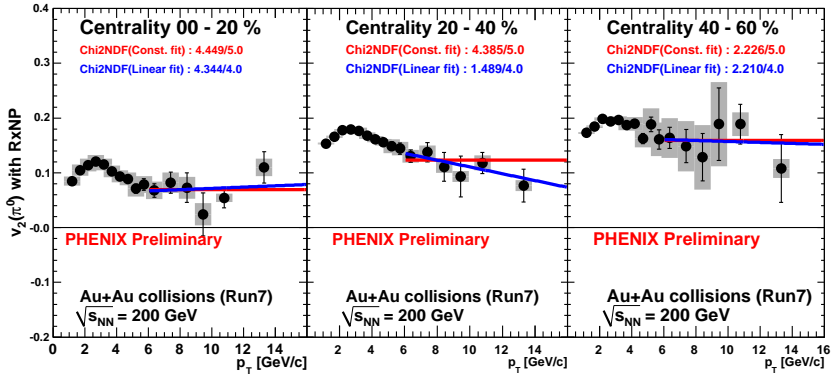}
    \caption{$v_{2}(\pi^{0})$ as a function of $p_{T}$ for each
      20~$\%$ centrality bin.
      Red and blue lines show constant and linear function,
      respectively. 
      Red and blue lines are fitted to data from 6~GeV/$c$.}
    \label{Fig:v2fitting}
\end{figure}

Reaction plane detector (RxNP) was installed in RHIC-Year~2007 and
reaction plane determination accuracy has been improved by a factor of
two as compared to that in RHIC-Year~2004.
Fig.~\ref{Fig:v2fitting} shows $v_{2}(\pi^{0})$ as a function of
$p_{T}$ for each 20~$\%$ centrality bin.
This detector enabled us to measure $v_{2}(\pi^{0})$ up to 14~GeV/$c$.
Measured $v_{2}(\pi^{0})$ are non-zero for all centrality bins.
Additionally, two assumed function is fitted to this data.
One is linear function ($f(v_{2})=a\cdot p_{T}+b$) and the other is constant
value ($f(v_{2})=c$).
At centrality 0-20~$\%$, values of $\chi^2$/NDF for constant and linear 
function are $4.45/5$ and $4.34/4$, respectively.
At centrality 20-40~$\%$, values of $\chi^2$/NDF for constant and linear 
function are $4.39/5$ and $1.49/4$, respectively.
At centrality 40-60~$\%$, values of $\chi^2$/NDF for constant and linear 
function are $2.23/5$ and $2.21/4$, respectively.
These results indicate that the values of $v_{2}(\pi^{0})$ in most
central and peripheral collisions tend to be constant while in
mid-central collisions tend to decrease.   


\section{Summary and outlook}
Study of path length and azimuthal angle dependence of R$_{AA}$ with
the new reaction plane detector has been started.  
The nuclear modififation factor, $R_{AA}$(5$<$$p_{T}$$<$8~GeV/$c$) as a
function of azimuthal angle has been compared to theoretical models.    
Even though the models are in good agreement with the (azimuthally
integrated) $R_{AA}$, so far they could not reproduce the measured
azimuthal angle dependence.   
The measurement of $v_{2}(\pi^{0})$ has now been extended to 14~GeV/$c$.  
For the most central (0-20$\%$) and peripheral (40-60$\%$) collisions,
the elliptic flow tends to be constant (instead of decreasing
monotonically with $p_{T}$), while for mid-central collisions
(20-40$\%$) we observe a decrease with $p_{T}$.

With the new data we will be able to measure the dependence of
$R_{AA}$ on azimuthal angle up to higher transverse momenta than ever
before. 
We can also estimate the path length by measuring the azimuthal angle
from reaction plane and mapping it into the shape of the participant
region, which can be calculated by Glauber model for each impact 
parameter (centrality).

%
%
%

%
\end{document}